\documentstyle[11pt,aas2pp4]{article}
\newcommand{\etal}{{{ et al.}}~}
\newcommand{\eg}{{{e.g.,}}~}
\newcommand{\ie}{{{i.e.,}}~}
\newcommand{\kms}{{{km s$^{-1}$}}~}
\newcommand{\kmsc}{{{km s$^{-1},$}}~}
\newcommand{\kmsp}{{{km s$^{-1}.$}}~}

\newcommand{\amin}{{{$^\prime$}}~}

\begin{document}

\title{A Cluster Merger and the Origin of the Extended Radio Emission in Abell 3667}

\vspace{1.in}
\author{KURT ROETTIGER}
\affil{Department of Physics and Astronomy\\University of Missouri-Columbia\\Columbia, MO 65211\\email: kroett@hades.physics.missouri.edu}
\author{JACK O. BURNS}
\affil{Office of Research and Dept. of Physics and Astronomy\\University of Missouri-Columbia\\Columbia, MO 65211\\email: burnsj@missouri.edu}
\author{JAMES M. STONE}
\affil{Department of Astronomy \\ University of Maryland\\ College Park, MD 20742-2421
\\jstone@astro.umd.edu}

\vspace{.5in}

\begin{center}{\bf Accepted for Publication in ApJ}
\end{center}

% Notice that each of these authors has alternate affiliations, which
% are identified by the \altaffilmark after each name.  The actual alternate
% affiliation information is typeset in footnotes at the bottom of the
% first page, and the text itself is specified in \altaffiltext commands.
% There is a separate \altaffiltext for each alternate affiliation
% indicated above.

%\altaffiltext{1}{NAS/NRC Associate}
%Visiting Astronomer, Cerro Tololo Inter-American Observatory. 
%CTIO is operated by AURA, Inc.\ under contract to the National Science
%Foundation.} 
%\altaffiltext{2}{Society of Fellows, Harvard University.} 
%\altaffiltext{3}{present address: Center for Astrophysics,
%    60 Garden Street, Cambridge, MA 02138}
%\altaffiltext{4}{Visiting Programmer, Space Telescope Science Institute}
%\altaffiltext{5}{Patron, Alonso's Bar and Grill}

% The abstract environment prints out the receipt and acceptance dates
% if they are relevant for the journal style.  For the aasms style, they
% will print out as horizontal rules for the editorial staff to type
% on, so long as the author does not include \received and \accepted
% commands.  This should not be done, since \received and \accepted dates
% are not known to the author.

\begin{abstract}

We present a numerical model for the extended steep-spectrum radio sources and the
elongated X-ray structure in A3667 based
on new 3-dimensional MHD/N-body simulations.
The X-ray and optical analyses of A3667 indicate that it has undergone
a recent subcluster merger event. We believe that the Mpc-scale radio
sources identified in A3667 are also a consequence of the merger.
Our previous numerical simulations show that mergers often produce large-scale
shocks and turbulence capable of both magnetic field amplification and 
in-situ reacceleration of relativistic particles. Our model
suggests that these radio structures, separated by $\sim$2.6$h^{-1}_{100}$ Mpc,
are in fact causally linked via a slightly off-axis merger that occurred
nearly in the plane of the sky approximately 1 Gyr ago with a subcluster
having a total mass equal to $\sim$20\% of the primary cluster.

\end{abstract}


\section { INTRODUCTION}

Abell 3667 (see Fig. \ref{a3667}) at z=0.055 (Sodr\'e \etal 1992) is a massive ($>$ 10$^{15}$ M$_\odot$; 
Sodr\'e \etal 1992; Knopp \etal 1996), X-ray luminous ($\sim$10$^{45}h^{-2}$ erg s$^{-1}$; Knopp \etal 1996)  
cluster of galaxies which exhibits a variety of the
observational signatures of a recent cluster merger (Roettiger \etal 1996).
Here, we present a self-consistent numerical MHD/N-body model suggesting that the most unique features of A3667, the Mpc-scale radio sources (R\"ottgering \etal 1997), are similarly a signature of a recent merger. 

Characterized variously as a ``halo" or a ``relic", we believe that the extended radio emission in A3667 represents
a stage in the evolution of a radio halo. Radio halos are generally described as large ($>$0.5 Mpc),
amorphous steep-spectrum structures which do not appear to be associated with any particular galaxy and,
as such, are believed to be intrinsic to the cluster itself (Hanisch 1980). By contrast, a
radio relic was once associated with a particular AGN that has turned-off and possibly
drifted away leaving the radio source to diffuse and fade away. 
It is possible that radio relics supply the seed relativistic particles necessary to produce
radio halos, but the connection is unclear at this time and we do not address this issue here.

Theoretically, the difficulty in understanding radio halos is related to their large spatial extent.
If the relativistic electrons responsible for the radio synchrotron emission originated
from a single source (\eg AGN), they would be required to diffuse over Mpc-scales during their
relatively short radiative lifetimes (of order 10$^8$ yrs) requiring velocities well in excess of the
limiting Alfv\'en speed. (Holman, Ionson \& Scott (1979) present an alternate view.) This has led several 
researchers to suggest that radio halos require
a means by which relativistic particles can be accelerated in-situ, such as via galactic wakes (Roland 1981) 
or shocks and
MHD turbulence (\eg Eilek \& Henriksen 1984) generated during a cluster merger (DeYoung 1992; Tribble 1993b). Whereas galactic wakes,
appear to be insufficient to power the radio halo (Goldman  \& Repheali 1991; DeYoung 1992), cluster mergers,
having a kinetic energy comparable to the total thermal energy of the cluster, are more than
sufficient to power the radio source (Tribble 1993b; Burns \etal 1994; Burns 1998). Mergers may also overcome
part of the electron transport problem by supplying bulk flows ($>$1500 \kms; Roettiger \etal 1997) capable of transporting the relativistic
electrons over large distances during their radiative lifetime (R\"ottgering \etal 1994).

Observationally, mergers have been linked to
radio halos by both a correlation with substructure (X-ray and optical) and by
an anticorrelation with cooling flows (Edge, Stewart \& Fabian 1992). For example, Coma, A2255 and A2256 all contain radio halos of one type
or another and exhibit substructure that is indicative of recent dynamical evolution. 
It has also been suggested that cooling flows (\eg see Fabian 1994 for a review of cooling flows) will be
disrupted by massive mergers (McGlynn \& Fabian 1984; Edge \etal 1992; Burns \etal 1997).  Gom\'ez \etal (1999), using
numerical simulations, have shown that mergers can disrupt cooling flows under various initial conditions.  
None of the above mentioned clusters contains a cooling flow, nor for that matter does A3667.

The purpose of this paper is to demonstrate using fully 3-dimensional, numerical MHD/N-body simulations
that the X-ray, optical and radio observations of A3667 are consistent with a single merger model in which
shocks formed during the merger provide sites for diffusive shock acceleration of cosmic ray (CR) electrons (see
Longair (1994), and references therein, for a review of particle acceleration in shocks). We do not attempt to model the CR acceleration,
diffusion, advection, and subsequent ``aging" self-consistently with the MHD. When done in detail, this is a difficult
problem and can be computationally expensive. Recently,  Jun \& Jones (1998) and Jones, Ryu \& Engler (1998) presented
a promising approach  to this problem (applied to supernova remnants) which we will look to incorporate into future work. At
this time, we present a model in which we introduce a population of CR electrons based on the shock characteristics during
an epoch believed to be representative of A3667's current dynamical state, as inferred from the X-ray and optical
data. The CR diffusion, advection, and
aging are modeled by assuming that the  MHD properties (\ie shock strength, structure and velocity, bulk flows etc.) 
do not evolve significantly during the radiative lifetime of the particles ($\sim$10$^8$ years) and 
that particle reacceleration behind the shock is unimportant. The modeling presented here is similar to that of
Tribble (1991, 1993a, 1994) but with two important exceptions. His models were purely static while ours are based
on the self-consistent evolution of shocks and magnetic fields within the merger model, and we are attempting
to specifically address the characteristics of the extended radio emission in A3667. 
Although others have suggested that A3667 has recently undergone a merger (Knopp \etal 1996), and still others have
suggested a possible shock origin for the extended radio emission  (R\"ottgering \etal 1997; Ensslin \etal 1998), this is the first study that presents a quantitative numerical model which attempts to simultaneously explain the radio, optical and X-ray morphology of this cluster within the context of a single merger.

Section \ref{obs} is a review of
recent observations and evidence for substructure in A3667. 
In \S\ref{num}, we discuss our numerical method. Our initial conditions
are presented in \S\ref{init}.  Section \ref{model} summarizes our model constraints. We discuss
the radio halo model and its limitations in \S\ref{radio}. Finally, we summarize
our results in \S\ref{summary}.  The model is scaled assuming H$_\circ$=70 \kms Mpc$^{-1}$. The
observational data (see \S\ref{obs}) is parameterized by h=H$_\circ$/100.

\section{A REVIEW OF THE OBSERVATIONS}
\label{obs}
\subsection{The X-ray Data}
\label{xdat}
Knopp \etal (1996) provide a detailed analysis of the {\it ROSAT} X-ray data which we summarize here.
The X-ray surface brightness (XSB; contours, Fig. \ref{a3667}) is seen to be elongated from the southeast (SE) to 
the northwest (NW). The ellipticity is
greatest in the cluster core and decreases with increasing radius. There is an isophotal twisting of
$\sim$15$^\circ$ from the inner to outer cluster. The XSB exhibits an extension to the NW of
the core while it is seen to be flattened to the SE. This results in a significant centroid shift (Mohr \etal 1995).
A3667 does not appear to be isothermal. Knopp \etal find a mean temperature of 6.3$\pm^{0.5}_{0.6}$ keV. However,
the core is found to be hotter (9$\pm^{4.0}_{2.0}$ keV) than the mean at the 82\% confidence level.
Markevitch \etal (1998), using {\it ASCA} data, find a mean temperature of 7.0$\pm0.6$ keV. 
Using the {\it ROSAT}-derived XSB, Knopp \etal find the total X-ray luminosity within 0.80$h^{-1}$ Mpc
to be L$_x$=6.74$\pm$0.09$\times$10$^{44}$$h^{-2}$ erg s$^{-1}$.
They estimate the gas mass within 0.80$h^{-1}$ Mpc to be 8.9$\pm$0.9$\times$10$^{12}$ $h^{-5/2}$ M$_\odot$ while
the total dynamical mass within the same radius (assuming hydrostatic equilibrium,
which is questionable in a recent merger) is 3.5$\pm^{0.3}_{0.4}\times$10$^{14}$ $h^{-1}$ M$_\odot$, giving a relatively low baryon fraction of 2.5$h^{-3/2}$\%.  The total mass within 3.0 $h^{-1}$ Mpc is estimated to be 1.4 $\pm0.2\times$ 10$^{15}$ $h^{-1}$  M$_\odot$.

\subsection{The Optical Data}
\label{odat}
Sodr\'e \etal (1992) provide 203 galaxy positions (128 redshifts) in the vicinity of A3667. The mean cluster redshift
is 16557$\pm$117 \kmsc and the velocity dispersion is 1223$\pm^{87}_{72}$ \kmsp Fadda \etal (1996), using a somewhat
larger galaxy sample and an analysis of substructure, find a global velocity dispersion of 971$\pm^{62}_{47}$ although
the velocity dispersion in the core appears to be much higher. Using the global velocity dispersion,
and the {\it ASCA} mean temperature, one calculates $\beta_{spec}$=0.84. This value is somewhat
discrepant from the value of $\beta_{fit}$=0.535 (Knopp \etal 1996). Numerical simulations have linked the so-called ``$\beta$-discrepancy" (Sarazin 1988) to recent merger activity (Navarro, Frenk \& White 1995; Roettiger \etal 1996).  Sodr\'e \etal further note two 
spatially distinct
clumps within the galaxy distribution. One clump is centered on the brightest D galaxy and the XSB peak while the other
is associated with the second brightest D galaxy located $\sim$16.6\amin to the NW (but still within the observed XSB distribution). Figure \ref{a3667g} shows the positions of galaxies having redshifts supplied by Sodr\'e \etal and the
results of the $\delta$-test for substructure (Dressler \& Shectman 1988). The circles, marking individual galaxy positions, are scaled in radius according
to their respective $\delta$ parameter. The value of $\delta$ indicates the degree to which a galaxy and it's 10 nearest neighbors
deviate from the global velocity properties. In Figure \ref{a3667g}, we have filled the circles corresponding to galaxies having the
lowest $\delta$ parameters, thereby accentuating the location of the two groups noted by Sodr\'e \etal. The degree
of substructure demonstrated by the $\delta$-test gives some indication as to why the observed range of velocity dispersions,
and therefore cluster virial masses, is so large. Sodr\'e \etal (1992) estimate a total mass of 2.6$\times$10$^{15}$$h^{-1}$ M$_\odot$
for the entire cluster while Biviano \etal (1993) find 1.6$\times$10$^{15}$$h^{-1}$ M$_\odot$ within 0.75$h^{-1}$ Mpc.
Both values are larger than the X-ray derived values (see \S\ref{xdat}).

\subsection{The Radio Data}
\label{rdat}
There are many radio point sources located within the spatial extent of A3667, but for our purposes, we are concerned
only with the large extended radio sources located to the SE and NW of the X-ray core (See Fig. \ref{a3667}, grayscale). The uniqueness of the extended radio sources in A3667 have been known for 
many years (Schillizzi \& McAdam 1975; Goss \etal 1982;
Jones \& McAdam (1992)). Here, we summarize the most detailed analysis to date, R\"ottgering \etal (1997). 
Approximately 1.7$h^{-1}$ Mpc NW of the cluster's X-ray core (and beyond the extent of the XSB), 
is an extremely large ($\sim$1.4$h^{-1}$ Mpc) arc-shaped source that does not appear to be associated with any
particular galaxy in the cluster. Along the NW rim of this source, there is a sharp edge to the synchrotron distribution.
The source becomes more diffuse to the SE. The spectral index appears to be relatively flat ($\alpha$=0.5;
S$_\nu \propto \nu^{-\alpha}$) along the NW 
edge and exhibits a gradient toward the SE steepening to $\alpha$=1.5. Overall, $\alpha\sim$1.1. 

To the SE of the X-ray core there are two large, linear sources situated perpendicular to the X-ray major axis
(parallel to each other). The largest and furthest SE extends $\sim$0.7$h^{-1}$ Mpc and is $\sim$0.7$h^{-1}$ Mpc
from the core. The source closest to the core has been identified by Hunstead and Wieringa (1998) as a narrow-angle tailed source (NAT) associated with galaxy 157 from Sodr\'e \etal (1992) and is therefore not of interest to this study. The NW and SE halo sources are separated by nearly 2.6$h^{-1}$ Mpc. Both sources are located near or beyond the  extent of {\it ROSAT}-observed X-ray emission.

\section{NUMERICAL METHOD}
\label{num}

The numerical method is the same as that used in Roettiger, Stone \& Burns (1998b).
The ICM and magnetic fields are evolved using
ZEUS (Stone \& Norman 1992a,b), an Eulerian, finite-difference
code which solves, self-consistently, the equations of ideal MHD.
The numerical evolution of the magnetic field components is
performed by the constrained transport (CT) algorithm (Evans \& Hawley 1988) which guarantees
preservation of the divergence-free constraint at all times. The
method of characteristics (MOC) is used for computing the electromotive
force (Hawley \& Stone 1995). An extensive series of MHD test problems have demonstrated that
the MOC-CT method provides for the accurate evolution of all modes of MHD
wave families (Stone \etal 1992). We employ outflow boundary conditions
on the MHD.

The collisionless dark matter is evolved using an N-body code based on a standard 
particle-mesh algorithm (PM, Hockney \& Eastwood 1988). The particles and gas
are evolved on the same grid using the same time step. The time step
is determined by applying the Courant condition simultaneously to both the dark 
matter and the magnetohydrodynamics.
The only interaction between the collisionless particles and the gas is 
gravitational. Since we are modeling an isolated
region, the boundary conditions for Poisson's equation are determined by a multipole expansion (Jackson 1975)
of the total mass distribution (dark matter and gas) contained within the computational grid.  
Particles that leave the grid are
lost to the simulation. Typically, less than a few percent of the particles leave the grid.

The hybrid ZEUS/PM code was parallelized using the Message Passing Interface (MPI; Gropp, Lusk \& Skjellum 1994).
These simulations were run on the Cray T3E in the Earth and Space Data Computing Division
of the NASA Goddard Space Flight Center.

The simulation whose results are presented here had an effective resolution scaled to 18 kpc, or more
significantly, there are 16 zones per primary cluster core radius. The computation was performed on a grid
having dimensions 512 $\times$ 256$^2$. The grid is uniform from zone 100 to 412 along the merger
axis and from zone 90 to 166 perpendicular to the merger axis. Outside of this central region, we
gradually increase the zone size by $\sim$3\% from one zone to the next.

\section {INITIAL CONDITIONS}
\label{init}
\subsection{Dark Matter and Gas}

Our initial conditions are similar to those used in our previous studies
(\eg Roettiger, Burns \& Loken 1993; Roettiger, Loken \& Burns 1997; Roettiger, Stone \& Mushotzky 1997, 1998a).
We begin with two clusters whose gas distributions are  consistent with
observations of relaxed systems. Our cluster dark matter distributions may be less consistent with observations
in that they do not have the central density cusp implied by recent
strong lensing observations and numerical simulations of large-scale structure formation (Navarro, Frenk, \& White 1997). 
We do not believe that this difference will significantly alter the merger dynamics.
The mass distributions in our simulations are based on the lowered isothermal King model described
in Binney \& Tremaine (1987). The lowered isothermal King model is a family of
mass distributions characterized by the quantity $\psi/\sigma^2$ which
essentially defines the concentration of matter. As $\psi/\sigma^2$
increases, the core radius ($r_c$) decreases with respect to the
tidal radius ($r_t$). We have chosen a model with  $\psi/\sigma^2$=12 in 
which we have truncated the density distribution at 15$r_c$. Near the half-mass
radius, the total mass density follows a power-law distribution, $\rho 
\sim$ r$^{-\alpha}$, where $\alpha \sim$ 2.6. This model is consistent with
mass distributions produced by cosmological N-body simulations which
show $\alpha \sim$ 2.4 in a high density universe ($\Omega$=1) and $\sim$2.9
in a low density universe ($\Omega$=0.2) (Crone, Evrard, \& Richstone 1994).
Initially the gas distribution (ICM) is in hydrostatic equilibrium within the gravitational
potential defined by both the gas and dark matter components. The gas distribution
is isothermal within the central 6$r_c$. At larger radii, the temperature drops gradually.
The density peaks are initially separated by 6.5 Mpc. Each cluster was given a small initial
velocity in order to speed up the merger process and thus conserve computational resources.
The final impact velocity ($\sim$2750 \kms) is not  affected significantly by the initial velocity
which has components of $\sim$275 \kms parallel to the line of centers and $\sim$30 \kms perpendicular to the line of 
centers, resulting in a slightly off-axis merger. The physical scaling of the pre-merger clusters
can be found in Table 1.

\subsection{Magnetic Field}
\label{maginit}

The only galaxy clusters certain to have large-scale or intrinsic magnetic fields are those with radio halos, 
and these sources are rare (Hanisch 1982). Of those containing halos, Coma has been studied in the most detail
(\eg Jaffe 1977; Kim \etal 1990 among others).
Deiss \etal (1997) show that the synchrotron halo in Coma, after subtracting point sources, traces the X-ray 
surface brightness distribution indicating
that the magnetic pressure gradient is similar to that of the thermal pressure. Of course Coma is
only one cluster and may not be representative of halos in general.   In addition
to the spatial distribution of the magnetic pressure, we must also be concerned with the
spectral power distribution. Cluster magnetic fields are believed to exist
as flux ropes that are tangled on a variety of physical scales (\eg Ruzmaikin \etal 1989).  Observations
of polarization in cluster radio sources indicate tangling occurs on scales ranging from a kiloparsec 
(\eg Feretti \& Giovannini 1997) to tens or even hundreds of kiloparsecs 
(Kim \etal 1990; Kim, Kronberg, \& Tribble 1991; Feretti \etal 1995) .
 Studies of cluster radio halos using a variety of methods (see Miley 1980)
indicate that typical mean field strengths are of order 1 $\mu$G (\eg Coma, Kim \etal 1990; Ensslin \& Biermann 1998; 
A2255, Burns \etal 1995; A2256, R\"ottgering \etal 1994; Bagchi, Pislar \&
Lima Neto (1998)).

In order to model a random field tangled on a variety of scales, we begin by defining a 
magnetic vector potential, A(k)=A$_\circ$k$^{-\delta}$, where the amplitudes (A$_\circ$) for each Cartesian
coordinate are drawn from
a Gaussian distribution. The vector potential is then transformed via a 3-dimensional
Fast Fourier Transform (FFT) into physical space where it is scaled spatially by the
gas density distribution. Assuming a uniform spherical collapse (likely a gross
over-simplification, \eg Evrard 1990;  Bryan \etal 1994; among many others) and flux freezing, it can be shown that the magnetic field will scale as $\rho_{gas}^{2/3}$ or B$\sim r^{-1.7}$.
Finally, we initialize a tangled divergence-free magnetic field from {\bf A} via ${\bf B}=\nabla \times {\bf A}$.  
This results in {\bf B$^2$} $\propto$ k$^{2(1-\delta)}$. Here we adopt $\delta=5/3$. Unfortunately, neither
observation or theory are capable of constraining our choice of power-law index.
We have scaled the magnetic pressure such that the mean field is a dynamically insignificant 0.3 $\mu$G within 2$r_c$.
Due to numerical resolution limitations, the minimum resolved magnetic field scale is $\sim$4 zones or nearly 80 kpc.

\section{THE MODEL}
\label{model}

Based on our numerical model and limited survey of merger parameter space (cluster mass ratios, impact parameters, relative
core densities, etc.), we believe that A3667 has undergone a merger
approximately 1 Gyr ago with a subcluster having a mass 20\% of the primary cluster. The merger
was slightly off-axis having an impact parameter of $<$0.5$r_c$.  The merger occurred largely in
the plane of the sky with the subcluster moving from the SE to the NW passing on the southern side of the more massive or primary
cluster.  Figure \ref{simtemp}a shows the synthetic XSB (dashed contours) overlaid onto the projected,
emission-weighted temperature (solid contours)  at
the epoch determined to most closely resemble A3667. The sharp edges in the temperature distribution indicate the locations
of shocks which we associate with sites of particle acceleration for the purposes of modeling the radio emission (See \S\ref{radio}).
Figure \ref{simtemp}b is a comparison of the simulated and {\it ASCA}-observed temperature profiles (Markevitch \etal 1998).
Figure \ref{simim}a shows the synthetic XSB (solid contours) overlaid on the simulated synchrotron emission (grayscale; see \S\ref{radio}). The
$+$ symbols indicate the location of dark matter/galaxy centroids. The primary cluster is largely coincident with the XSB 
peak while the subcluster remnant is located to the NW. The future evolution of this system would be similar to that of A754 as described in detail by
Roettiger \etal (1998a).

In producing our model of A3667, we have made some effort to match as many of A3667's observational characteristics
(see \S\ref{obs}, Figs. \ref{a3667} and \ref{a3667g}) as  closely as possible, while keeping in mind that our primary concern for this work is the merger-induced
shock structure which, as our previous simulations show (\eg Roettiger \etal 1997) are a rather ubiquitous
feature of major merger events. Regarding the XSB distribution, we are able to reproduce
the overall elongation which defines the merger axis, the  isophotal twisting which indicates a slightly
off-axis merger, and the  steepness of the XSB distribution to the SE relative to the NW.
We believe that the subcluster first impacted from the SE moving to the NW causing a compression of the X-ray
emitting gas on the SE side. As the subcluster dark matter remnant exits the primary cluster core, it draws out the
ICM creating the NW extension. With regard to the galaxy distribution, we are able to reproduce
the relative positions of the galaxy concentrations (dark matter particles in our simulations, $+$'s in Fig. \ref{simim}a) with respect to
the XSB and the radio sources. Below, we describe radio source modeling in greater detail below.

\section {DETAILS OF THE RADIO HALO MODEL}
\label{radio}

\subsection{Basic Assumptions}

There are several simplifying
assumptions inherent in this model.  First, we assume
that all particle acceleration occurs at the shocks. Once the
particles leave the region of the shock, there is no further acceleration. 
Second, we assume a simple diffusive shock model in which the strength of the shock (as measured by the density compression) 
determines the power-law slope of the injected particle energy spectrum (\eg Drury 1983; Longair 1994). The electrons are assumed
to be accelerated within a single zone. That is, their scattering length is small compared to a computational
zone, and the time required to accelerate them is small compared to a time step. Third, we assume that the basic shock structure (strength, shape,
velocity, etc.) does not change significantly during the radiative lifetime of the relativistic
electrons. Finally, we assume that some small fixed fraction of the thermal particle flux is accelerated to its
equilibrium distribution as it passes through the shock.

\subsection{Method}
\label{meth}

Having chosen the epoch most representative of A3667 based on the X-ray and optical data, we begin modeling the radio
emission by identifying the locations of the shocks. In ZEUS, an artificial viscosity is used
to smooth shocks over 4-5 zones. We locate the shock by calculating the gas compression ratio
($r=\rho_2/\rho_1$) at two points separated by 5 zones throughout the computational volume. The compression ratio can be related directly
to the power-law index ($\gamma$) of the injected relativistic electron energy spectrum, $N(E)\propto E ^{-\gamma}$
such that $\gamma=$${3.0} \over{(r-1)}$$+1$. Once identified, each point along the shock is assigned a value of
$\gamma$ based on the local strength of the shock. For the most part, the shock strength is relatively uniform
across the face of the shock, however it does weaken somewhat near its edges. 
The shock structure present within the plane of the merger can be seen in Fig. \ref{shock}. Note that there
are multiple shocks to the SE of the X-ray core each presenting a potential site for
particle acceleration.

The
next step is to associate each zone behind (or upstream from) the shock with the nearest point on the shock along the fluid
flow lines.
We then assume that the particles in a given zone behind the shock are derived from that location on the shock
and that their distance ($d$) from the shock is directly proportional to the velocity of the shock relative to the
bulk fluid flow ($V_s$), and the time since acceleration ($t_a$) such that $d=\kappa V_s t_a$. The
scaling factor $\kappa$ is used to parameterize the diffusion/advection rate. If the electrons diffuse
freely in straight-line motion, unaffected by the magnetic field or turbulence, $\kappa$=1. 
This would represent the limiting case in which the magnetic field is
weak and/or aligned parallel to the flow and accelerated particles are simply deposited in
the wake of the shock. Of course, the situation is almost certainly more complicated.
At the other extreme (strong, tangle field), Tao (1995) suggests that the maximum inhibition
of electron diffusion by magnetic fields is of order a factor of ten. That is, the effective path length
of an electron traveling through a tangled magnetic field environment will be
at most ten times longer than the straight-line path, (\ie $\kappa$=0.1). Here, we adopt the
weak field/high diffusion limit where $\kappa\sim$1.0.

In order to quantify the rate at which particles diffuse/advect from the shock in the absence
of a magnetic field (\ie $V_s$), we have incorporated a passive-scalar field into ZEUS. The passive-scalar
is a dynamically insignificant quantity (in that sense, it is likely similar to the radio plasma) which
is advected with the thermal gas density. In this implementation, we inject the scalar field at the locations
of shocks and simply allow it to evolve. As expected, the passive-scalar field (or relativistic electrons) is seen to trail the shocks as they
propagate through the ambient medium. This is consistent with observations of shocks within the solar system (Kennel \etal 1986).
By evolving the scalar field over a limited time interval near the A3667 epoch, we can estimate
the effective shock velocity for each point along the shock front. In this manner, we find effective
shock velocities of 700-1000 \kms while the SE shocks have an effective velocity of $\sim$250 \kmsp
In fact, the shock nearest the X-ray core is effectively stationary.
This analysis shows little evidence of mixing between the relativistic and thermal plasmas behind the NW shock. 
Of course this result could be strongly
resolution dependent. It may also depend on the initial conditions. The shocks in this model, particularly to
the NW, propagate into a largely non-turbulent, homogeneous medium which is certainly not the case.

\subsection{Synchrotron Aging, Emissivity, and Spectral Index Distribution}

We use the formalism of Myers \& Spangler (1985) to age our synchrotron spectrum. Their work is
based on the models of Kardashev (1962) and Pacholczyk (1970) and Jaffe \&  Perola (1973). Here,
we specifically employ the modification introduced by Jaffe \& Perola which assumes that the electrons become isotropized on timescales short compared to their radiative lifetimes and therefore can be
represented by a time-averaged distribution of pitch angles. Adiabatic loses are
not included in this model.  Based on our analysis of the simulation dynamics and the R\"ottgering \etal (1997) analysis
showing the radio source to be greatly underpressurized with respect to the thermal plasma, we believe that expansion losses
can be neglected.

Given the initial electron energy spectrum power-law index ($\gamma$), the magnetic field strength ($B$), and the
time since acceleration, we can determine the spectral index and relative intensity of emission between 1.4 GHz and 4.9 GHz. The emissivity
within a given zone is then simply, $\epsilon_{\nu} \propto (B sin(\theta))^{\alpha+1} \nu^{-\alpha}$, where
$\theta$ is the angle between the magnetic field and the observer's line-of-sight. The Myers \& Spangler model does
not account for varying magnetic field strengths. Therefore, the field used to age the particles is the mean field 
experienced during the electron's lifetime (\ie the mean magnetic field along the line connecting the particle's 
current location and it's presumed point of origin on the shock).  
The magnetic field used to calculate the emissivity within a given zone is simply the field within that zone.
Having calculated the synchrotron emissivity within each volume, we perform a line-of-sight integration through the emitting
volume to produce the surface brightness image (Fig. \ref{simim}a).

The rate at which the particles age (or equivalently, their synchrotron spectrum steepens) is determined by the 
strength of the magnetic field traversed by the particles. Although the simulation was run with
a specified field strength, since it was dynamically insignificant, we can treat it as a free parameter (within
a reasonable range of values) for purposes of modeling the radio spectral index distribution.
Examining Fig. 3 of R\"ottgering \etal (1997), we see that the spectral index of the NW source
steepens to $\sim$1.5 at about 0.7 Mpc (H$_\circ$=70 \kms Mpc$^{-1}$) upstream from the shock. Assuming the
Jaffe-Perola model and high diffusion ($\kappa=1$; see \S\ref{meth}), we can reproduce the distribution
of spectral index if the mean magnetic field within the region of the source is $\sim$0.6 $\mu$G.
If we restrict the rate of diffusion (\ie $\kappa<1$), the magnetic field can be decreased accordingly because
particles will take longer to diffuse/advect a distance of 0.7 Mpc and will have aged spectrally in the mean time.
It is not possible to use this method to constrain the field strength in the SE radio sources. We do
not have spectral index data, and the shocks have not moved significantly during the particle's lifetimes.
Therefore, we scale the global field according to our analysis of the NW source. Assuming we had
the correct initial magnetic pressure distribution (see \S\ref{maginit}), the mean field in the region of the SE radio
source is $\sim$1.5 $\mu$G.
The greater field strength results from the SE source being closer to the X-ray core than the NW source  (remember B$\sim r^{-1.7}$, initially), and from the greater compression (and amplification) of the field on the SE side of the cluster
as a result of the merger. Again, if we have overestimated the electron diffusion/advection rate, as is likely, these
field estimates will decrease.

\subsection{Limitations of the Model}
\label{limits}

Of course the model has its limitations which are owed primarily to the large number of free and not necessarily independent parameters such as relative cluster masses, the shape of their initial mass distribution,
impact parameter, initial velocities, magnetic field distribution, projection with respect to the observer, etc.
It is computationally expensive and physically impossible to explore all of parameter space. Consequently,
we present the best fit to the current data given our limited survey of the many free parameters.
Even so, the largest discrepancies between our model and the observations appear to be
the strength of the NW shock and the location of the most northwesterly of the SE shocks. The NW shock in our model, based on the observed spectral index distribution, appears to be too weak. If the spectral index along
the NW rim of the NW radio source ($\alpha$=0.5) is correct, we need a shock with a density compression
near 4 (\ie a strong shock). At the chosen epoch, it is typically less than 3 which
produces a somewhat steeper spectrum ($\alpha\sim$0.95). 
This may imply that we need to consider an earlier epoch since the shock weakens as it expands outward. The shock could also be strengthened if the impact velocity of the subcluster where
greater (\ie more massive clusters or a greater initial velocity), or if the exiting subcluster
remnant were interacting with higher density gas at large radii or with material falling into the cluster from a large-scale filament.
This highlights another limitation of the model. These numerical simulations are non-cosmological. We
do not attempt to model the external cluster environment such as filaments or spherical accretion.
The structure of galaxy clusters at large radii ($>$1.5$h^{-1}$ Mpc) is uncertain from an observational standpoint, and,
consequently, our initial conditions become potentially less accurate at larger radii. 

As mentioned above, our simulation produces essentially two shocks located
SE of the X-ray core whereas only one extended radio source has been observed that does not appear to be associated with
a particular galaxy. Both shocks appear to have a morphology, strength and evolution consistent with producing
the SE radio source. That is, they are both smaller, more linear and slower moving than the NW shock while still
being strong enough to produce significant particle acceleration. In terms of its location with respect
to the XSB, the SE shock furthest from the core is more consistent with the observed radio source. However, its
properties suffer from the same sensitivity to initial conditions as we described above regarding the NW shock. That said, 
the important point is that the single merger does produce
shocks on either side of the cluster core which are capable of generating the observed radio morphology. Why some shocks
within a cluster may produce extended radio sources and others do not will likely depend on the specific details of the local
magnetic field structure and relativistic particle populations. These are the same issues that likely  determine
why some clusters have radio halos and others do not (see \S\ref{why}).

\subsection{Why are there not more Halos of this type?}
\label{why}

If the large shocks are ubiquitous features of major merger events and mergers appear to be relatively common,
why don't we see more radio halos of this type? Specifically, according to our models, A754 is in a
stage of evolution similar to (though
somewhat earlier) than A3667, but it does not have the same type of radio halo (Roettiger \etal 1998a). Andernach \etal (1988)
claims the existence of a radio halo in A754. However, higher resolution images indicate that this source is composed of
three discrete sources, one of which is a NAT (Owen \& Ledlow 1997). The other two are likely background objects.
In any case, there does not appear to be a halo in A754 that is comparable to A3667. There are several possible explanations for this apparent
inconsistency. First, the shock structures, although long-lived relative to the canonical radio source age,
are still relatively short-lived compared to the period between major mergers (2-4 Gyrs; Edge \etal 1992). Second, although
the shocks may represent a necessary condition of halo formation, by themselves, they are not sufficient.
Certainly, a large-scale magnetic field is required and possibly a pre-existing population of relativistic
electrons that are reaccelerated by the shock. At this time, very little is known about either the large-scale
distribution of magnetic field or relativistic particles as a general property of clusters. Some clusters may simply
not have them or may have them in insufficient quantities. Third, cluster richness appears to be a factor since
only the richest appear to have halos (Hanisch 1982). Richer (\ie more massive) clusters have deeper gravitational potentials which provide
for more violent, higher velocity mergers and stronger shocks.  They also have a more extended, higher pressure ICM,
and more galaxies which may supply seed relativistic electrons and magnetic fields (Okoye \& Onuora 1997). 
Finally, projection effects may also play a role. 

\section {SUMMARY}
\label{summary}

We have presented a plausible model for the extended radio emission in A3667.
In this model, A3667 has undergone a recent ($\sim$1 Gyr) merger event with a subcluster having
a total mass of 20\% of the primary cluster. We believe that the subcluster impacted slightly off-axis while
 moving from the SE to the NW. The merger generates multiple shocks which
provide sites for diffusive shock acceleration of relativistic electrons. The spatial extent of the
radio sources is determined by the shock dimensions as well as the assumed strength and structure of the 
magnetic field which ultimately determines the rate at which the relativistic  particle age spectrally
and the rate at which they diffuse/advect from the shock. This model is capable of reproducing several features of
the basic X-ray morphology including the twisted isophotes (although the twisting is more
extreme in our model), the steep X-ray surface brightness gradient to the SE (indicating a compression
of the ICM), and the X-ray extension to the NW (indicating that the ICM has been drawn out of the cluster
by the subcluster remnant). By modeling the spectral aging of the electron distribution, we estimate
a mean magnetic field of order less than 1 $\mu$G.

This model also explains many features of the A3667 radio morphology including: 1) the locations
of the NW and SE radio sources with respect to the X-ray surface brightness (these are the were the shocks are located)
, 2) the sharp edge of the NW radio source (the site of particle acceleration), 
3) the steepening of the  spectral index toward the southeast in the NW source  (particles age as 
they move away from the shock), 4) the relative sizes of the radio sources (the NW shock has
a much greater lateral extent and is moving much faster than the SE shocks which are virtually stationary 
at this epoch), and, 5) the shapes of the radio components (the SE shocks are more linear than the NW shock).

Although we have been able to reproduce the basic radio morphology, there are limitations to this
model. The NW shock, in our model, is not strong enough to produce the flat ($\alpha=0.5$) spectrum
observed on the leading edge of the source. This shock could be strengthened if there is infall
along an external filament (which we do not model) as suggested by Ensslin \etal (1998). The SE shocks
are stronger than the NW shock largely because of residual infall from the subcluster which impinges
on gas expelled from the cluster core by oscillations in the gravitational potential.  It is also important to note 
that we do not evolve
the relativistic particles self-consistently within the MHD simulation. Modeling the diffusion of
particles from the shock is very complicated. For simplicity, we have assumed a weak field (high
diffusion) limit in which particles are immediately left behind the shock. 
The true situation is undoubtedly more complex. Future work will entail evolving
the relativistic particles in a more self-consistent manner as a second-fluid (\eg Jun \& Jones 1998;
Jones \etal 1998).

At this time we can only speculate as to why this type of radio source is so rare.
It is most likely that very specific initial conditions are required that may not exist in
all clusters. It is possible that not all clusters contain a cluster-wide magnetic field or a seed population of relativistic electrons.
It is also likely that the relatively short radiative lifetime of the relativistic particles combined with
the short lifetime of the shocks contributes to the rarity of these sources.

We thank the Earth and Space Data Computing Division at the NASA Goddard Space Flight Center
for use of the Cray T3E supercomputer.  We acknowledge support
from NSF grant AST-9896039. We also thank H. R\"ottgering for use of Fig. \ref{a3667}.

{\bf Note to be added in proof:} Markevitch, Sarazin, \& Vikhlinin (1998) recently produced a temperature
map of A3667 using {\it ASCA} data. Our model is largely consistent with this map over the limited
extent of the X-ray emission. The greatest inconsistency is the very hot gas SE of the X-ray
core which is not apparent in the data. This would seem to indicate that our subcluster may
be penetrating too deep into the cluster.
\newpage

\section*{ REFERENCES}
 \everypar=
   {\hangafter=1 \hangindent=.5in}

{
Andernach, H., Tie, H., Sievers, A., Reuter, H.-P., Junkes, N., \& Wielebinski, R. 1988, A\&AS, 73, 265

Bagchi, J., Pislar, V., \& Lima Neto, G. B. 1998, MNRAS, submitted (astro-ph/9803020)

Binney, J, \& Tremaine, S. 1987, Galactic Dynamics, (Princeton: Princeton University Press)

Biviano, A., Girardi, M., Giuricin, G., Mardirossian, F., \& Mezzetti, M. 1993, ApJ, 411, L13

Bryan, G. L., Klypin, A., Loken, C., Norman, M. L., \& Burns, J. O. 1994, ApJ, 437, 5L

Burns, J. O., Roettiger, K., Pinkney, J., Perley, R. A., Owen, F. N., \& Voges, W. 1995, ApJ, 446, 583

Burns, J. O., Roettiger, K., Ledlow, M. \& Klypin, A. 1994, ApJ, 427, L87

Burns, J. O. \etal 1997, in Galactic \& Cluster Cooling Flows, ed. N. Soker (ASP:San Francisco), Vol. 115, 21

Burns, J. O. 1998, Science, 280, 345

Crone, M., Evrard, A., \& Richstone, D. 1994, ApJ, 434, 402

DeYoung, D. S. 1992, ApJ, 386, 464

Deiss, B. M., Reich, W., Lesch, H., \& Wielebinski, R. 1997, A\&A, 321, 55

Dressler, A. \& Shectman, S. A. 1988, AJ, 95, 985

Drury, L. 1983, Rep. Prog. Phys., 154, 973

Edge, A. C., Stewart, G. C., \& Fabian, A. C. 1992, MNRAS, 258, 177

Eilek, J. A. \& Henriksen, R. N. 1984, ApJ, 277, 820

Ensslin, T. A., \&  Biermann, P. L. 1998, A\&A, 330, 90

Ensslin, T. A., Biermann, P. L., Klein, U., \& Kohle, S. 1998, A\&A, 332, 395

Evrard, A. E. 1990, ApJ, 363, 349

Evans, C. R. \& Hawley, J. F. 1988, ApJ, 332, 659

Fabian, A. C. 1994, ARAA, 32, 277

Fadda, D., Girardi, M., Giuricin, G., Mardirossian, F., \& Mezzetti, M. 1996, ApJ, 473, 670

Feretti, L., Dallacasa, D., Giovannini, G., \& Tagliani 1995, A\&A 302, 680

Feretti, L., \& Giovannini, G. 1997, astro-ph/9709294

Goldman, I. \& Rephaeli, Y. 1991, ApJ, 380, 344

Gom\'ez, P. L, Loken, C., Roettiger, K. \& Burns, J. O. 1999, in preparation

Goss, W. M., Ekers, R., Kellern, D. J., \& Smith, R. M., 1982, MNRAS, 198, 259

Gropp, W., Lusk, E., \& Skjellum, A. 1994, Using MPI: Portable Parallel Programing with the Message-Passing Interface, (Cambridge: MIT Press)

Hanisch, R. J. 1980, AJ 85, 1565

Hanisch, R. J. 1982, A\&A 116, 137

Hawley, J. F., \& Stone, J. M. 1995, Comp. Phys. Comm., 89, 127

Hockney, R., \& Eastwood, J. 1988, Computer Simulation Using Particles, (Philadelphia: IOP)

Holman, G. D., Ionson, J. A., Scott, J. S. 1979, ApJ, 228, 576

Hunstead, R. W. \& Wieringa, M. 1998, private communication.

Jackson, J. D. 1975, Classical Electrodynamics, (New York: Wiley)

Jaffe, W. J. 1977, ApJ, 212, 1

Jaffe, W. J. \& Perola, G. C. 1973, A\&A, 26 423

Jones, P. A., \& McAdam, W. B., 1992, ApJS, 80, 137

Jones, T. W., Ryu, D., \& Engel, A. 1998, ApJ, astro-ph/9809081

Jun, B.-I. \& Jones, T. W. 1998, ApJ, astro-ph/9809082

Kardashev, N. S. 1962, Soviet Astr. --AJ, 6, 317

Kennel, C. F., Coroniti, F. V., Scarf, F. L., Livesey, W. A., Russell, C. T., Wenzel, E. J. \&
Scholer, M. 1986, J. Geophys. Res. 91, 11

Kim, K.-T., Kronberg P. P., Dewdney, P. E., Landecker, T. L., 1990, ApJ, 355, 29

Kim, K.-T., Kronberg, P. P., Tribble, P. 1991, ApJ, 379, 80

Knopp, G. P., Henry, J. P., \& Briel, U. G. 1996, ApJ, 472, 125

Longair, M. S. High Energy Astrophyics, (Cambridge, UK: Cambridge Univ. Press)

Markevitch, M., Forman, W. R., Sarazin, C. L., \& Vikhlinin, A. 1998, ApJ, 503, 77

Markevitch, M., Sarazin, C. L., \& Vikhlinin, A. 1998, ApJ Letters, submitted (astro-ph/9812005)

McGlynn, T. A. \& Fabian, A. C., 1984, ApJ, 208, 709

Miley, G, 1980, ARAA, 18, 165

Mohr, J. J., Evrard, A. E., Fabricant, D. G., \& Geller, M. J. 1995, ApJ, 447, 8

Myers, S. T., \& Spangler, S. R. 1985, ApJ, 291, 52

Navarro, J. F., Frenk, C. S., \& White, S. D. 1995, MNRAS, 275, 720

Navarro, J. F., Frenk, C. S., \& White, S. D. 1997, ApJ, 490, 493

Okoye, S. E. \& Onoura, L. I. 1997, MNRAS, 283, 1047

Owen, F. N,. \& Ledlow, M. J. 1997, ApJS, 108, 41

Pacholczyk, A. G. 1970, Radio Astrophysics, (San Francisco: Freeman)

R\"ottgering, H., Snellen, I., Miley, G., de Jong, J. P., Hanisch, R., \& Perley, R. A. 1994, ApJ, 436, 654

R\"ottgering, H., Wieringa, M. H., Hunstead, R. W., \& Ekers, R. D. 1997, MNRAS, 290, 577

Roettiger, K., Burns, J. O., \& Loken, C 1993, ApJ, 407, 53L

Roettiger, K., Burns, J. O., \& Loken, C 1996, ApJ, 473, 651

Roettiger, K., Loken, C., \& Burns, J. O. 1997, ApJS, 109, 307

Roettiger, K., Stone, J. M., \& Mushotzky, R. F. 1997, ApJ, 482, 588

Roettiger, K., Stone, J. M., \& Mushotzky, R. F. 1998a, ApJ, 493, 62

Roettiger, K.,  Stone, J. M., \& Burns, J. O. 1998b, ApJ, in press.

Roland, J. 1981, A\&A, 93, 407

Ruzmaikin, A., Sokoloff A., \& Shukurov, A. 1989, MNRAS, 241, 1

Sarazin, C. L. 1988 in X-ray Emissions from Clusters of Galaxies. Cambridge University
Press, Cambridge

Schilizzi R. T., \& McAdam, W. B 1975, Mem. R. Astro. Soc., 79, 1

Sodr\'e, L., Capelato, H. V., Steiner, J. E., Proust, D., Mazure, A. 1992, MNRAS, 259, 233

Stone, J. M. \& Norman, M. 1992a, ApJS, 80, 753

Stone, J. M. \& Norman, M. 1992b, ApJS, 80, 791

Stone, J. M., Hawley, J, F., Evans, C. R., \& Norman, M. L. 1992, ApJ, 388, 415

Tao, L. 1995, MNRAS, 275, 965

Tribble, P. C. 1991 MNRAS, 253, 147

Tribble, P. C. 1993a MNRAS, 261, 57

Tribble, P. C. 1993b MNRAS, 263, 31

Tribble, P. C. 1994 MNRAS, 269, 110

}
\newpage

\begin{center}{\bf FIGURE CAPTIONS}
\end{center}

\begin{figure}[htbp]
\centering \leavevmode
\epsfxsize=1.0\textwidth \epsfbox{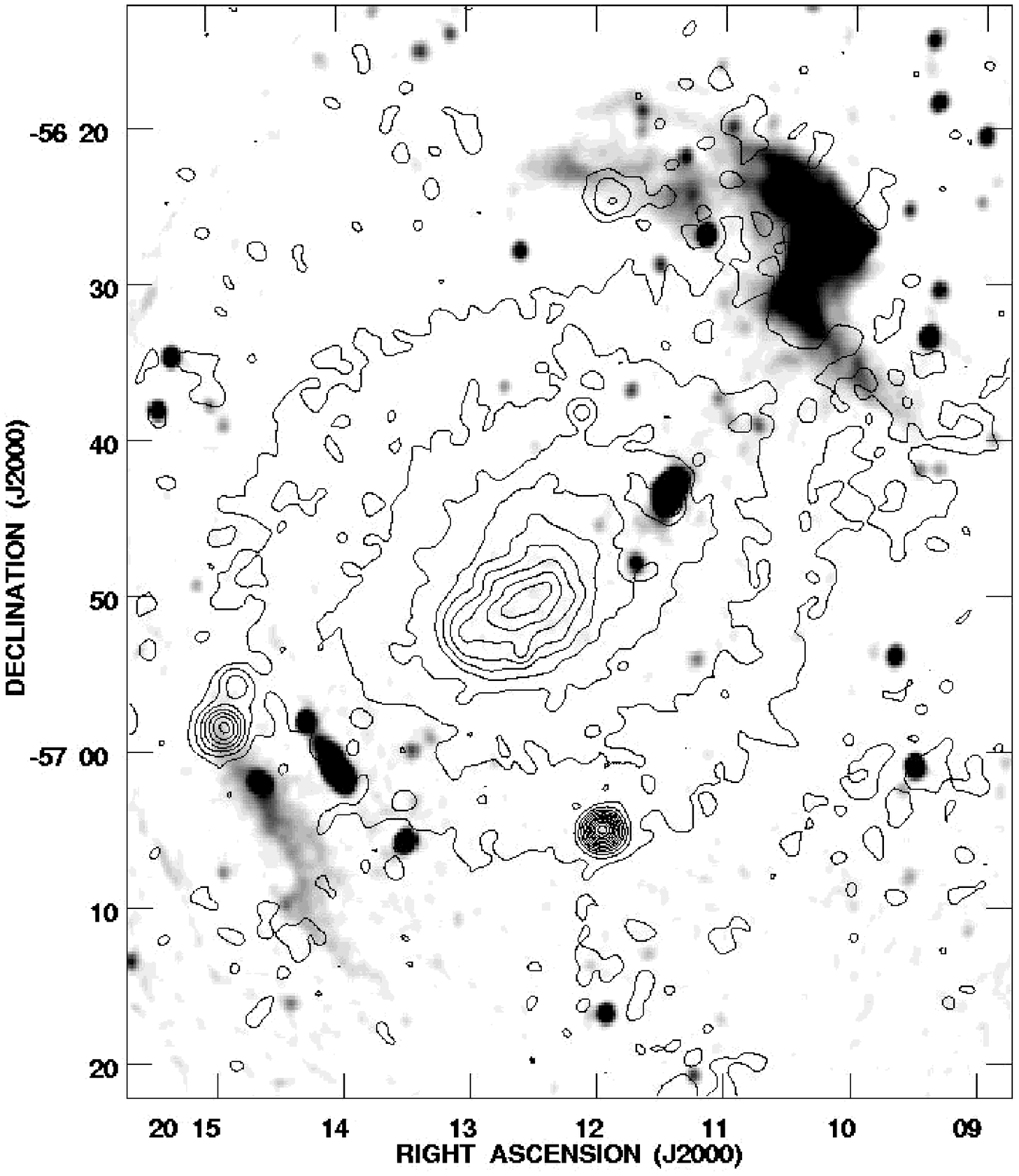}
\caption[]
{ }
\label{a3667}
\end{figure}

 {\bf Fig. \ref{a3667}} The {\it ROSAT} PSPC X-ray image (0.1-2.4 keV, contours)
overlaid on the grayscale of the wide-field, 843 MHz MOST image. The
X-ray contours correspond to 2, 8, 18, 32, 50, 72, 98, 128, 162, 200, and 242
times the background noise. The rms noise level of the radio image is 0.7 mJy beam$^{-1}$. (Figure courtesy H. R\"ottgering from R\"ottgering \etal 1997).

\begin{figure}[htbp]
\centering \leavevmode
\epsfxsize=0.9\textwidth \epsfbox{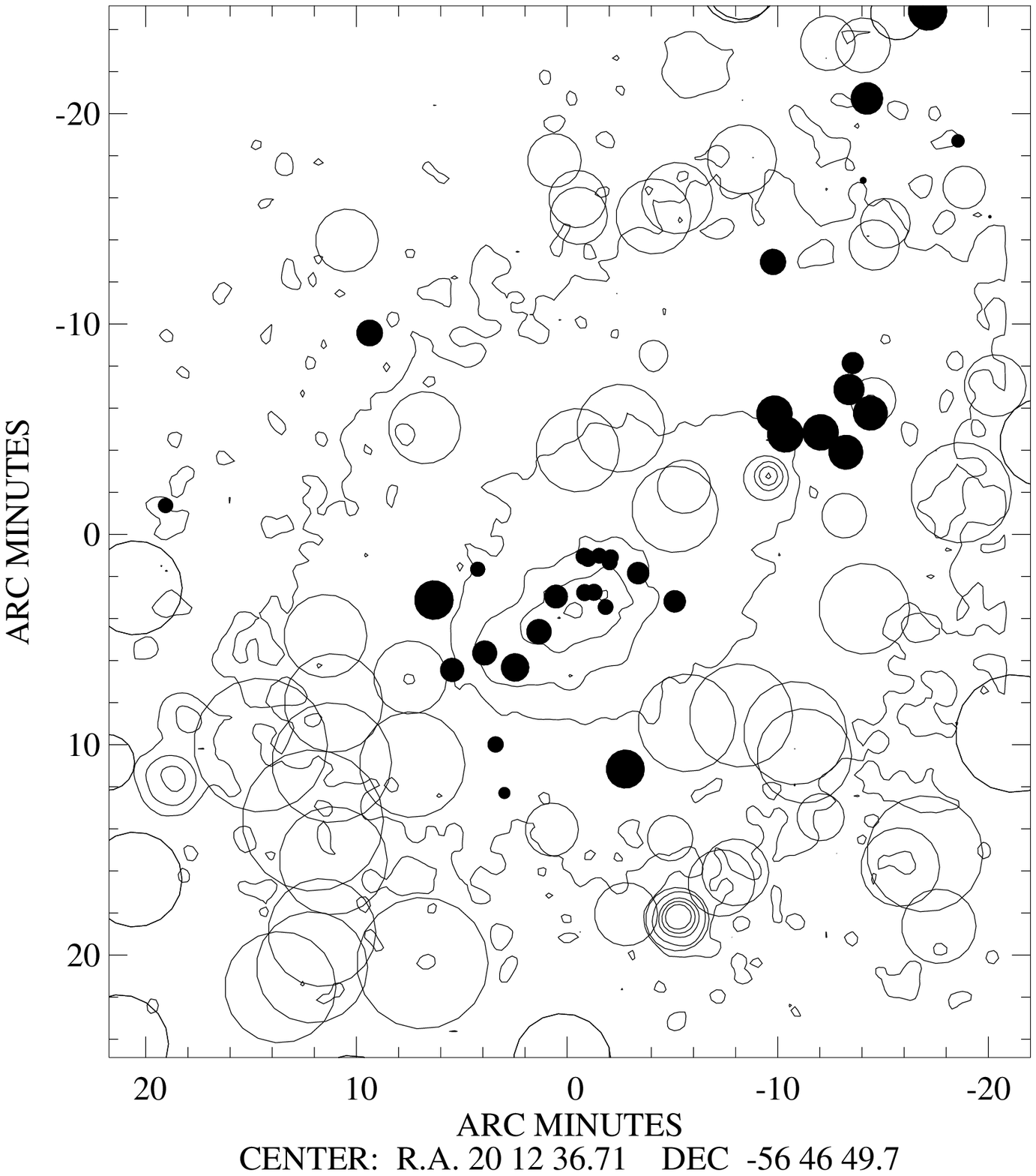}
\caption[]
{ }
\label{a3667g}
\end{figure}

{\bf Fig. \ref{a3667g}} Galaxy positions (both filled and open circles) 
overlaid on the {\it ROSAT} PSPC image (contours). The circle sizes are
proportional to the galaxy's $\delta$ statistic (Dressler \& Shectman 1988).
A large value of $\delta$ indicates a large local deviation in velocity mean
and dispersion from the global values. We have filled circles representing
local distributions having $\delta$ less than 20\% of the maximum.
Note the two regions associated with the XSB peak and to the NW of the
peak. Both of these are associated with D galaxies.

\begin{figure}[htbp]
\centering \leavevmode
\epsfxsize=1.0\textwidth \epsfbox{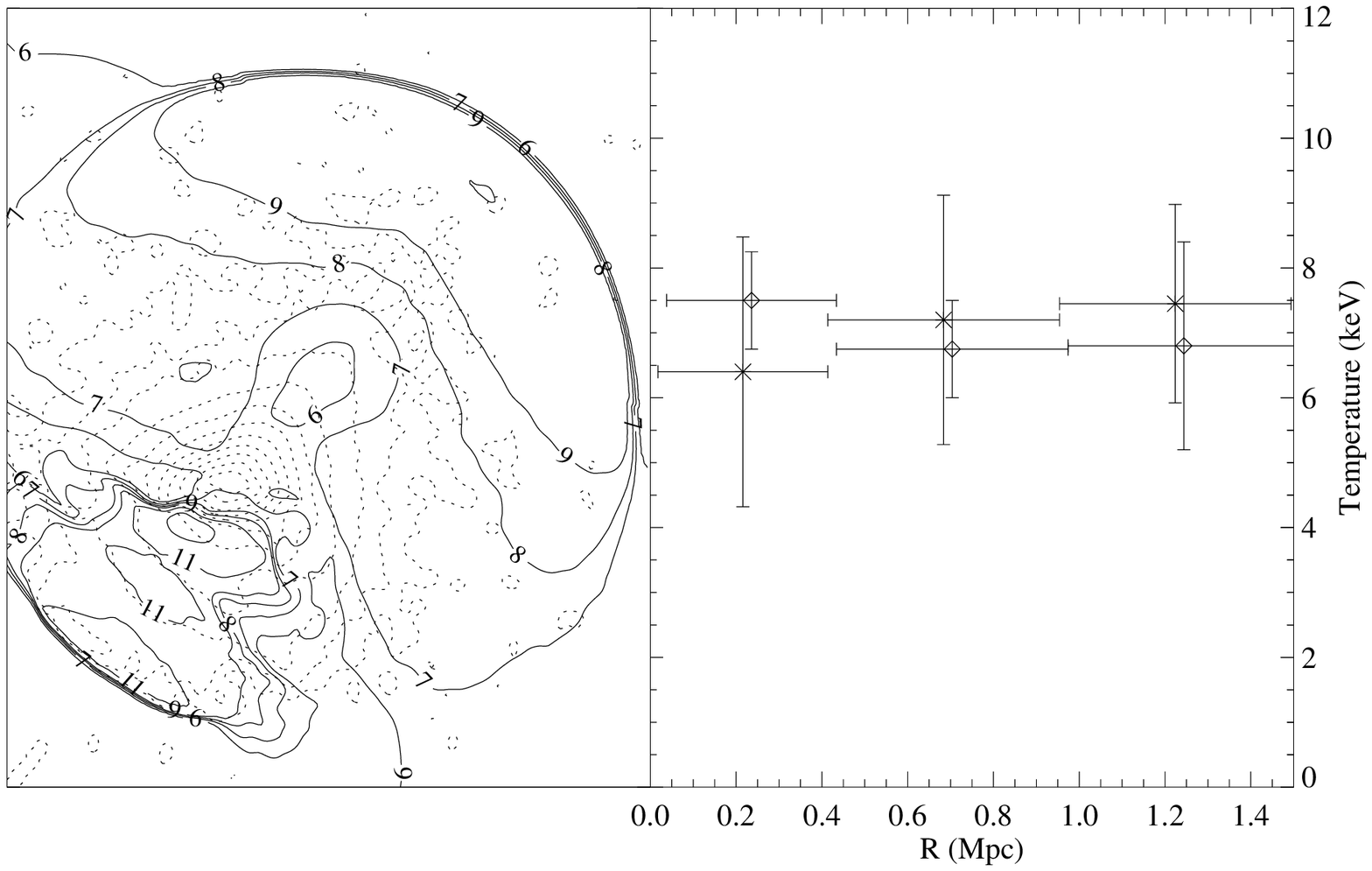}
\caption[]
{ }
\label{simtemp}
\end{figure}

{\bf Fig. \ref{simtemp}} a) The simulated X-ray surface brightness (dashed contours) and  projected emission-weighted temperature (solid contours).
We have added noise to the synthetic
X-ray image which has been contoured at the same levels (relative to peak) 
as the {\it ROSAT} image (see Fig. \ref{a3667}). 
The temperature contours are labeled with the corresponding temperature in
keV. The image is 3.15 $\times$ 3.85 Mpc. b) The {\it ASCA} temperature profile ( Markevitch \etal 1998, open diamonds) and the simulated temperature profile (asterisks) within the same
spatial bins (as indicated by the horizontal bar, assuming $H_\circ$=70 \kms Mpc$^{-1}$). The profiles have been slightly displaced for clarity. The vertical bars on the {\it ASCA} data indicate the uncertainty
in the temperature while the vertical bars on the simulated temperatures
represent the dispersion in temperature withing the volume. Note that
although the cluster is non-isothermal, it appears to be quite isothermal
in profile.

\begin{figure}[htbp]
\centering \leavevmode
\epsfxsize=1.0\textwidth \epsfbox{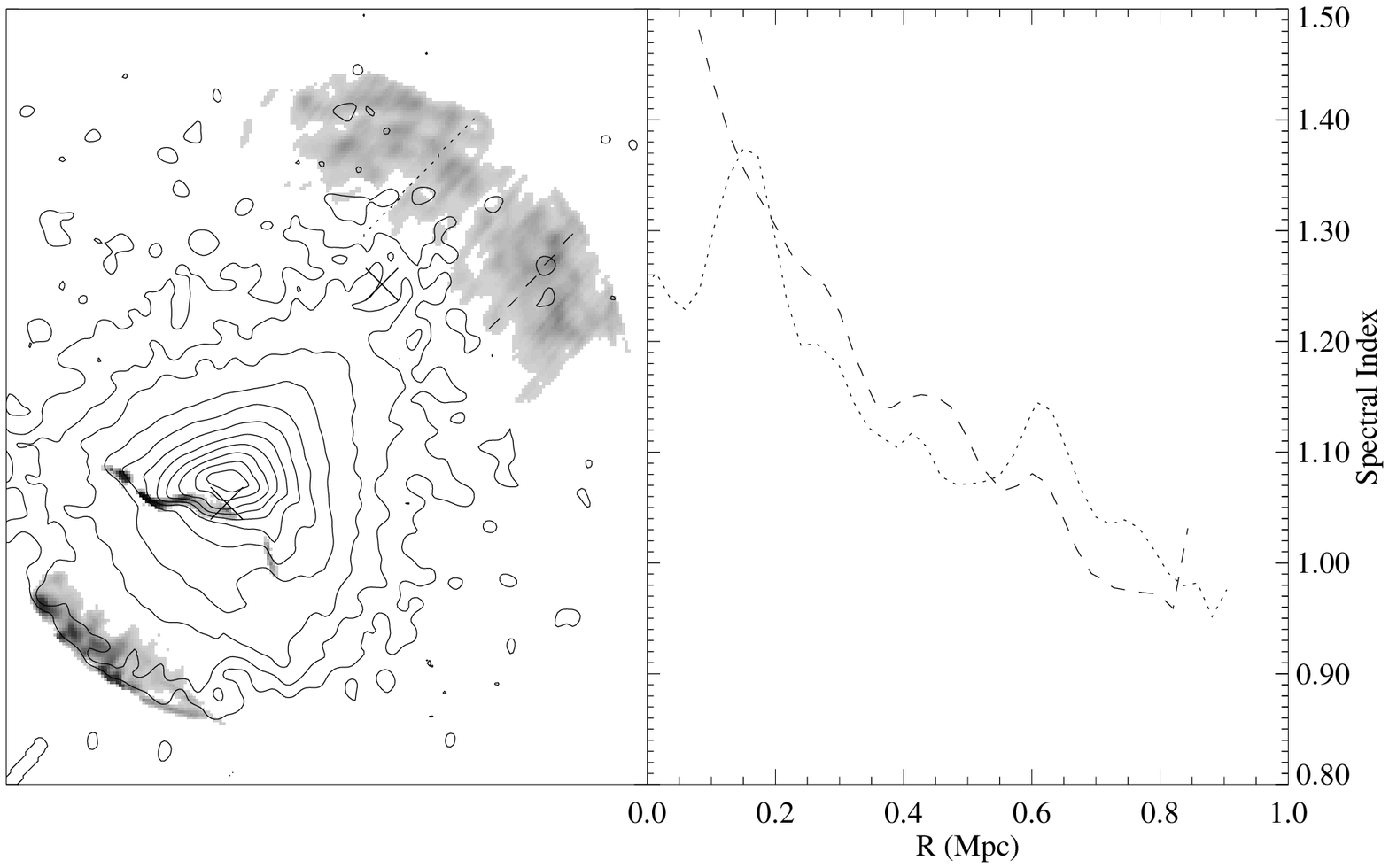}
\caption[]
{ }
\label{simim}
\end{figure}

{\bf Fig. \ref{simim}} a) The simulated A3667 X-ray surface brightness and radio data. Contours represent
the X-ray surface brightness. We have added noise to the synthetic
X-ray image which has been contoured at the same levels (relative to peak) 
as the {\it ROSAT} image (see Fig. \ref{a3667}). The grayscale represents 
the synthetic radio emission at 1.4 GHz. The $+$ symbols indicate the locations of peaks
in the dark matter/galaxy distribution. The dashed and dotted lines
traversing the synthetic radio emission to the NW (upper right) indicate
the location of the radio spectral index ($\alpha^{1.4}_{4.9}$) profiles displayed in (b).  The dashed
and dotted lines in (a) correspond to the dashed and dotted lines in (b).
The profiles extend from the southeast (lower left) to the northwest (upper right). The image is 3.15 $\times$ 3.85 Mpc.

\begin{figure}[htbp]
\centering \leavevmode
\epsfxsize=.6\textwidth \epsfbox{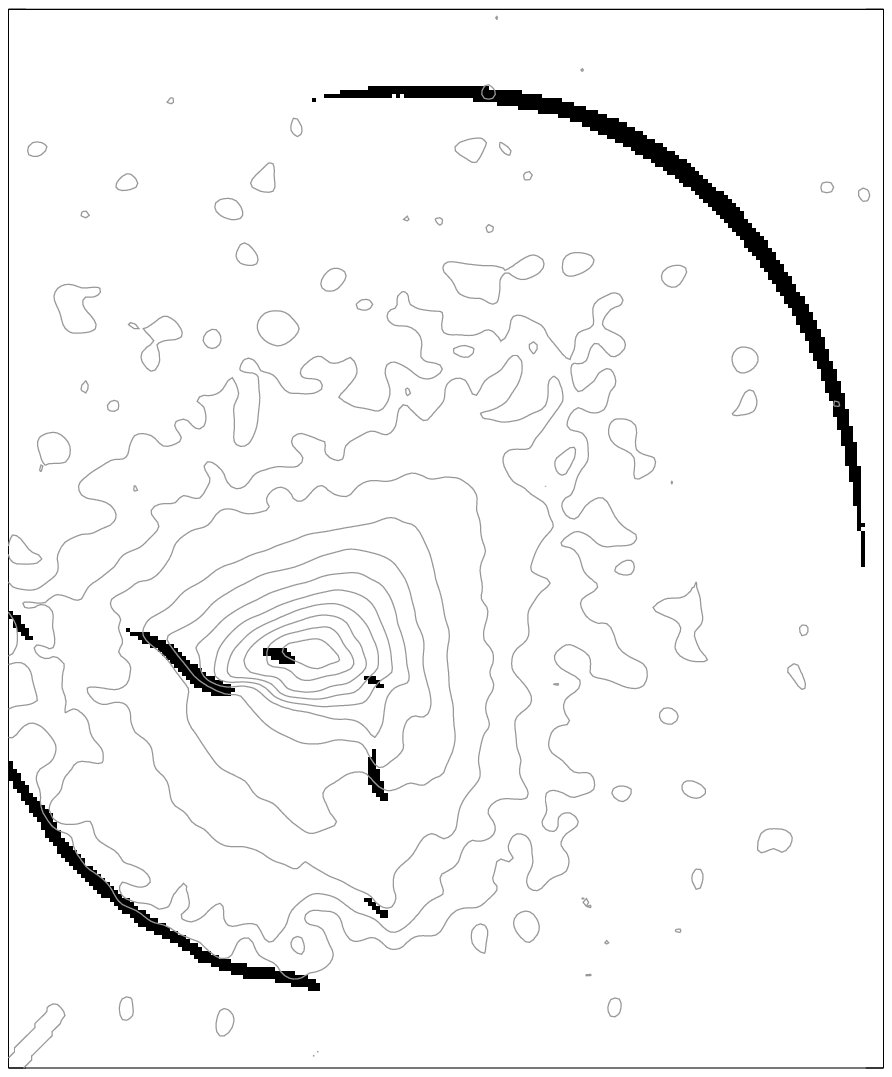}
\caption[]
{ }
\label{shock}
\end{figure}

{\bf Fig. \ref{shock}} The simulated X-ray surface brightness (contours, same
as in Figs. \ref{simtemp}a and \ref{simim}a) 
overlaid onto the underlying shock structure (solid shading) within
the plane of the merger. Regions with a density compression greater than two are shaded black. The image is 3.15 $\times$ 3.85 Mpc.

\onecolumn

\begin{table}
\label{tab1}
\begin{center}
\begin{tabular}{c c c c c c c} 
\multicolumn{7}{c}{Table 1. Initial Cluster Parameters}\\ \hline \hline

 \multicolumn{1}{c}{Cluster$^1$} & 
 \multicolumn{1}{c}{$M_{tot}^2$} &
 \multicolumn{1}{c}{$T_e^3$} &
 \multicolumn{1}{c}{$\sigma_v^4$} &
 \multicolumn{1}{c}{$r_c^5$} &
 \multicolumn{1}{c}{$f_g^6$} &
 \multicolumn{1}{c}{$v_{impact}^7$} \\

 \multicolumn{1}{c}{ ID } & 
 \multicolumn{1}{c}{ (10$^{14}$ M$_\odot$) } &
 \multicolumn{1}{c}{ (keV) } &
 \multicolumn{1}{c}{ (km s$^{-1}$) } &
 \multicolumn{1}{c}{ (kpc) } &
 \multicolumn{1}{c}{     } &
 \multicolumn{1}{c}{ (km s$^{-1}$)} \\ \hline

1 & 13.0 & 6.5 & 943  & 288 & 0.045 & 2750 \\
2 & 2.6 & 3.6 & 540  & 144 & 0.045  &  \\ \hline 
\end{tabular}
\end{center}
{ $^1$ Cluster ID. 
 $^2$ Total Mass R$<$3 Mpc.
 $^3$ Temperature.
 $^4$ 1-Dimensional Velocity Dispersion R$<$1.5 Mpc.
 $^5$ Core Radius. 
 $^6$ Global Gas Fraction, by mass. 
 $^7$ Impact Velocity (dark matter).}
\end{table}

%\onecolumn
\pagestyle{empty}
\newpage

\onecolumn

\onecolumn

\onecolumn

\end{document}